\documentclass[aps,pra,twocolumn,showpacs,superscriptaddress]{revtex4}
\usepackage{epsfig}
\usepackage{bm,amsmath,amssymb,latexsym,mathrsfs,graphicx,enumerate}
\usepackage[mathcal]{euscript}
\usepackage{hyperref}

\newcommand{\be}{\begin{equation}}
\newcommand{\ee}{\end{equation}}

\begin{document}

\title{The split-ring Josephson resonator as an artificial atom}

\author{J.-G.~Caputo$^{~1}$, I. Gabitov$^{~2}$  and A.I.~Maimistov$^{~3,4}$}

\affiliation{\normalsize \noindent
$^1$: Laboratoire de Math\'ematiques, INSA de Rouen, \\
BP 8, Avenue de l'Universite,
Saint-Etienne du Rouvray, 76801 France \\
E-mail: caputo@insa-rouen.fr \\
$^2$: Department of Mathematics,\\
University of Arizona, Tucson, AZ, 85704, USA \\
E-mail: gabitov@math.arizona.edu\\
$^3$: Department of Solid State Physics and Nanostructures, \\
Moscow Engineering Physics Institute, \\
Kashirskoe sh. 31, Moscow, 115409 Russia \\
E-mail: amaimistov@gmail.com \\
$^4$: Department of Physics and Technology of Nanostructures \\
 REC Bionanophysics,\\
Moscow Institute for Physics and Technology, \\
Institutskii lane 9, Dolgoprudny,
Moscow region, 141700 Russia }

\date{\today }

\begin{abstract}

Using the resistive-shunted-junction model we show that
a split-ring Josephson oscillator or radio-frequency SQUID in the
hysteretic regime is similar to an atomic system. It has a number of stationary
states that we characterize. Applying a short magnetic pulse we switch
the system from one state to another. These states can be detected
via the reflection of a small amplitude signal forming the base of a new
spectroscopy.

\end{abstract}

\maketitle

\section{Introduction}

Since the realization of lasers using atoms in the sixties, many researchers
have been trying to replicate this effect using quantum objects similar
to two-level systems. The basic elements to realize a laser are
\cite{encyclopedia_nl} (i) energy levels for the electrons, (ii) a pumping
mechanism to populate upper levels and (iii) an optical cavity to confine
photons enhancing the probability of interacting with excited electrons.
A Josephson junction between two 
superconductors is a macroscopic quantum system and Tilley \cite{tilley}
predicted that interconnected Josephson junctions could emit a coherent
radiation. 
Rogovin and Sculley \cite{rs76} also established 
a connection between Josephson junctions and two-level quantum systems.
The superradiance prediction was confirmed by Barbara et al \cite{bcsl99}
who showed that the power emitted by an array grows like
$N^2$ where $N$ is the number of active junctions in the array. 
See also the recent experiments by Ottaviani et al \cite{ocl09}
showing the synchronization of junctions in an array. Since the radiation 
emitted is in the Terahertz range where there are no solid-state sources, these
systems are being investigated as microwave sources (see for 
example \cite{ok07}).
Up to now however practical difficulties subsist and the 
powers emitted by such devices remain low.

From another point of view artificial materials (Metamaterials) have
been fabricated by embedding metal split-rings or rods into dielectric
materials. This way negative index materials have been 
fabricated\cite{Veselago:06}, \cite{Pendry:04}.
One can obtain artificial atoms with a high magnetic moment
and a nonlinear response to electromagnetic waves. 
Such an artificial atom would have many advantages over a real atom.
For atoms the dipole momentum is
very small and the interactions between them are small.
To realize interactions large density are necessary. Also in general the
excited state has an energy much larger than the energy of the 
dipole coupled with electromagnetic wave. 
Many candidates for artificial atoms have been proposed, most
of them width intrinsic nonlinearities. Among them we have
a diode \cite{Lapin:2003} , a Kerr material \cite{Zharov:2003}
or a laser amplifier \cite{Gabitov_Kennedy:2010}.
Another example is the Josephson junction
discussed above (see \cite{Maim:Gabi:10} for a review). The use of these devices in  metamaterials was
advocated by Lazarides\cite{Lazarides:06}
\cite{lt07} and by the authors 
\cite{Maim:Gabi:10}.
They introduced a split ring resonator with a Josephson junction
contact. In the Josephson community, this device is called an RF SQUID 
for Radio Frequency Superconducting Quantum Interference Device  \cite{Barone} 
\cite{Likharev}.
It is the elementary component of the arrays of junctions discussed above. In this work we will show that rather than the junction
itself, the RF SQUID can be considered as an artificial atom.

We will study the so-called hysteretic regime where the system has controlled
metastable states and show that one can switch from the ground state
to one of these excited states by applying a suitable flux pulse.
We show that a magnetic field can act on such a system in a similar way as 
an electric field acts on dipoles in atoms. We also show
that these states can be detected by examining the reflection 
coefficient of an electromagnetic wave incident on the device. This
is the base of spectroscopy. The 
main result of this study is to show that a split-ring Josephson
oscillator (RF SQUID) in the hysteretic regime behaves as an
artificial atom with discrete energy levels. It is the only device
that leads to such discrete levels as opposed to the systems mentioned above.
\\
The article is organized as such. In section 2 we derive
the model and analyze it in section 3.
In section 4 we characterize 
how to switch from one state to another and how the state can be detected. 
In the last section we study the scattering of an electromagnetic wave 
by a split ring Josephson resonator

\section{The model }

\begin{figure}
\centerline{\epsfig{file=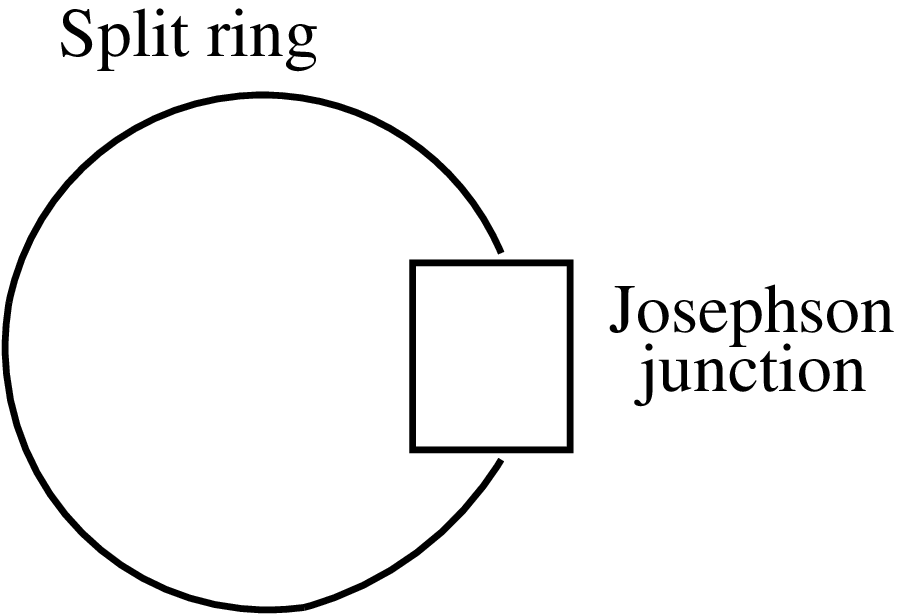,height=5cm,width=5cm,angle=0}
\epsfig{file=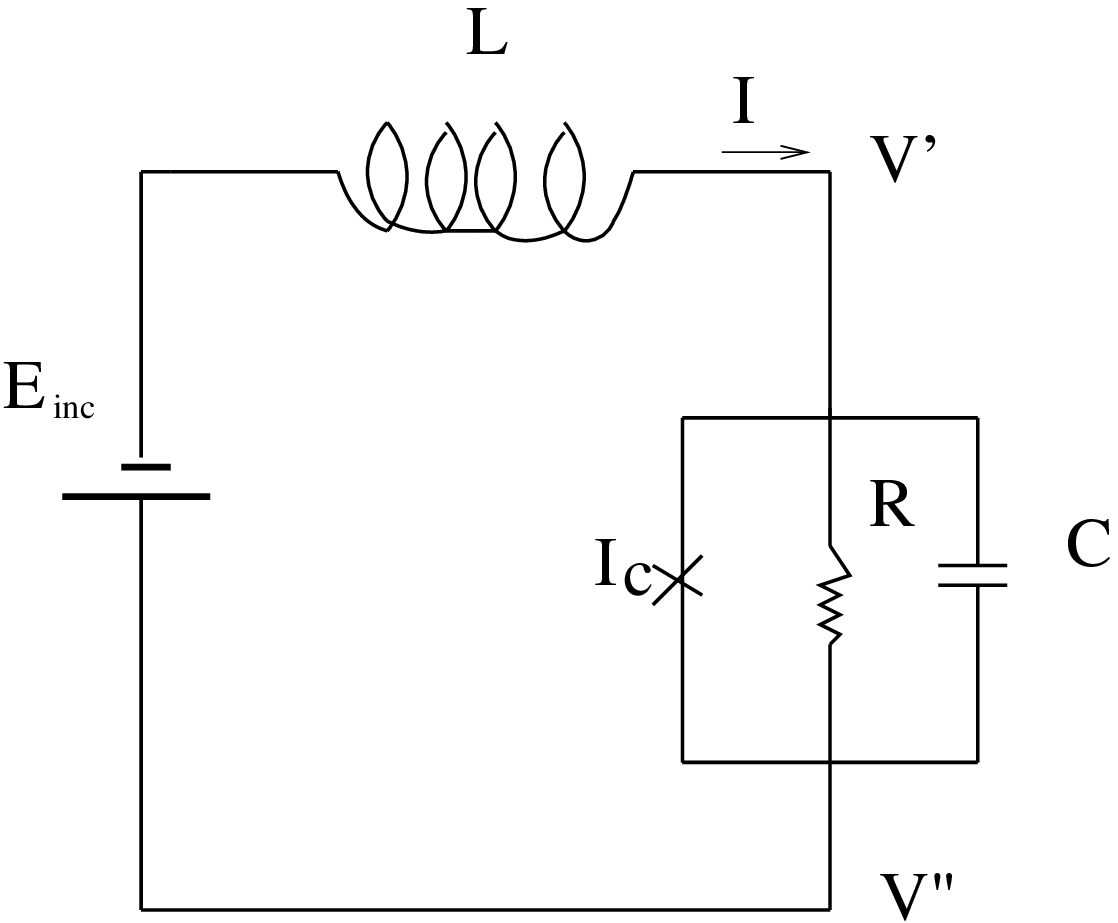,height=5cm,width=5cm,angle=0}}
\caption{A split ring resonator with an embedded Josephson junction 
(left panel). The right panel shows the equivalent circuit using 
the Resistively Shunted Junction model for the Josephson junction. }
\label{f1}
\end{figure}

The device we consider is shown in the left panel of
Fig. \ref{f1}. It is a split ring
resonator in which is embedded a Josephson junction. Practically
it can be made using a ring like strip of 
superconducting material where a small region was oxidized to 
make the junction. The right panel of Fig. \ref{f1} shows the
electric representation of the device, an inductance $L$ for
the strip and the Resistive Shunted Junction (RSJ) model for the
Josephson junction. The latter represents the junction as
a resistor $R$,  a capacity $C$ and the nonlinear element in parallel.
This last element is the sine coupling $I_c \sin \Phi/ \phi_0$ where
$\Phi$ is the magnetic flux and $\phi_0$ is 
the reduced flux quantum (see below).
In standard electronics the conjugate variables
are voltages and currents while in superconducting electronics 
they are the fluxes and charges where a flux is defined as
the time integral of a voltage. This is why we present the
derivation in detail here. The device is assumed to be operating
at low temperature so that losses are minimal and Josephson relations hold.

The Josephson equations describing the 
coupling of two superconductors
across a thin oxide layer are
\be\label{joseph}
V = {d \over dt} (\Phi'-\Phi") ~~, ~~ I = I_c \sin\left({\Phi'-\Phi" \over \phi_0}\right)~,\ee
where $V$ and $I$ are respectively the voltage and current across
the barrier, $\Phi',\Phi"$ are the 
macroscopic phases in the two superconductors,
$I_c$ is the critical current of the junction and
$\phi_0=\hbar/(2e)$ is the reduced flux quantum.
The right panel of Fig. \ref{f1}
shows the equivalent electric circuit of the whole system assuming
a Resistively Shunted Junction model \cite{Barone},\cite{Likharev} 
for the Josephson junction. 
We then define
$$\Phi' = \int_{-\infty}^{t} V'(\tau) d\tau,~~~
\Phi" = \int_{-\infty}^{t} V"(\tau) d\tau,$$
where $V', V"$ are the voltages on each side of the Josephson junction 
(see Fig. \ref{f1}). 
Kirchoff's law gives
\be\label{vpmvpp}
V'-V" = -L I_t + E_{e},\ee
where the subscript indicates time derivative and $E_{e}$ is
the electromotive force due to an electromagnetic pulse incident
on the ring. We neglect the resistance of this loop which we assume
to be made of superconducting material.
In terms of fluxes this relation is 
\be\label{fpmfpp}
\Phi'-\Phi" = -L I + \Phi_{e},\ee
where $\Phi_{e}$ is the incident flux.
Kirchoff law for node $V'$ gives
\be\label{kirchoffv}
I = I_c \sin({\Phi'-\Phi" \over \phi_0}) +{V'-V" \over R}
+ C (V'_t-V"_t),\ee
which in terms of the fluxes becomes
\be\label{kirchofff}
I = I_c \sin({\Phi'-\Phi" \over \phi_0}) +{\Phi'_t-\Phi"_t \over R}
+ C (\Phi'_{tt}-\Phi"_{tt}).\ee
We now introduce the phase difference
$\Phi\equiv \Phi'-\Phi"$ 
and combine equations (\ref{fpmfpp}) and (\ref{kirchofff}) to
obtain our final equation
\be\label{usrr_jj}
L\left [  C \Phi_{tt} + {\Phi_{t} \over R} + 
I_c \sin({\Phi\over \phi_0}) \right ] +\Phi=\Phi_{e}.\ee
The quantity in brackets is the current $I$ circulating in the loop.
To measure the importance of the $\sin$ term in this equation 
we introduce the dimension-less Josephson length as a ratio of 
the flux in the loop versus the flux quantum
\be\label{lamdaj}
\beta = { L I_c \over \phi_0 }.\ee
Time is normalized by the Thompson frequency, $t'= \omega_T t$ where
\be\label{thompson}
\omega_T = {1 \over \sqrt{L C}}.\ee
The fluxes are normalized by $\phi_0$ as $\Phi= \phi \phi_0$,
$\Phi_{e}=\phi_{e} \phi_0$
In the normalized time $t'$, dropping the ' for ease of writing we
get our final dimensionless equation
\be\label{srr_jj}
\phi_{tt} + \alpha \phi_{t} + \beta
\sin(\phi )+\phi=\phi_{e},\ee
where the damping parameter $\alpha$ is
\be\label{def_alpha}
\alpha = {\omega_T L \over R}.\ee

\section{Analysis of the model }

The ordinary differential equation (\ref{srr_jj}) can be written
as the 1st order system
\begin{eqnarray}
\phi_t  =\psi,\\
\psi_t  =-\alpha \psi -\beta \sin(\phi )-\phi + \phi_{e} .
\end{eqnarray}
The system has the fixed
points $(0,0)$ and $(\phi^*,0)$ where 
\be\label{fix_relation} - \beta\sin(\phi^* )- \phi^* + \phi_{e}= 0. \ee
A plot of the above relation indicates that for  $\beta > 4.34$
there are no additional fixed points. The fixed points can be approximated
for large $\beta$ using an asymptotic expansion. The equation 
(\ref{fix_relation}) can be written as
$$ \sin \phi^* + {1 \over \beta} (\phi^* -\phi_{e}) =0.$$
Writing the solution 
$$\phi^* = \phi_0 + {1 \over \beta} \phi_1 +
{1 \over \beta}^2 \phi_2 + \dots$$
we get
\be\label{approx_fix}
\phi^* = n \pi + {1 \over \beta} (-1)^n (\phi_{e}-n \pi) + \dots,\ee
where $n$ is an integer.

\begin{figure}
\centerline{\epsfig{file=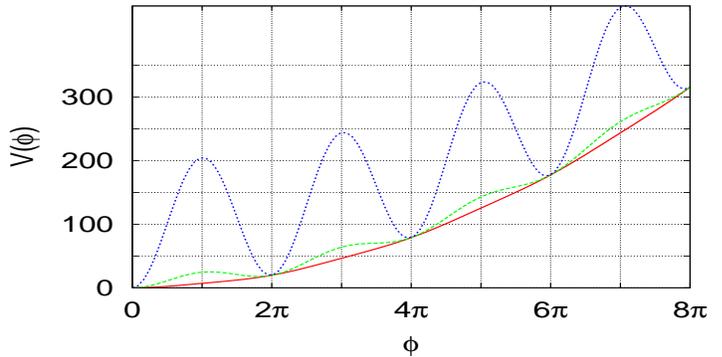,height=5cm,width=10 cm,angle=0}}
\caption{Potential energy $V(\phi)= 
\beta (1-\cos\phi)
+ {1\over 2} \phi^2$ with three different values of $\beta$, $\beta = 1, 9.76 
$ and $100$.}
\label{f2}
\end{figure}

In the absence of damping $\alpha=0$ and forcing $\phi_{e}=0$, 
the system is Hamiltonian with
\be\label{ham} 
H(\phi,\phi_t)= {1\over 2} \phi_t^2 + 
\beta (1-\cos\phi)
+ {1\over 2} \phi^2.\ee
The stable fixed points correspond to the minima of the potential
\be\label{pot}
V(\phi)= \beta (1-\cos\phi)
+ {1\over 2} \phi^2.\ee
Fig. \ref{f2} shows a plot of the potential $V(\phi)$ for
$\beta=1, 9.76$ and $100$. For $\beta=1$ shown as a continuous curve 
(red online) there is only one fixed point $\phi=0$.
For $\beta=9.76$ shown in dashed line (green online) there are three minima 
corresponding to stable fixed
points, $\phi=0, \pm \phi^*$ where $\phi^* \approx 2 \pi$. For
$\beta=100$ there are many stable fixed points.

\begin{figure}
\centerline{\epsfig{file=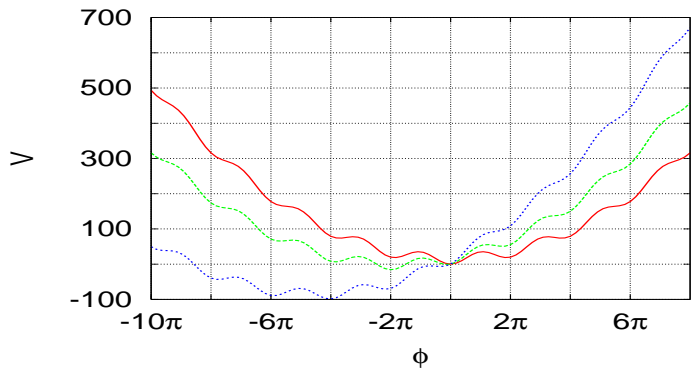,height=5cm,width=10 cm,angle=0}}
\caption{Potential energy $V(\phi)=
\beta (1-\cos\phi)
+ {1\over 2} \phi^2 + h \phi$ for three different values of the static
incident flux $\phi_e  = 0$ in continuous line (red online), $\phi_e =1.8 \pi$
in dashed line (green online) and $\phi_e =4.5 \pi$ in short dashed line 
(blue online).}
\label{f2a}
\end{figure}

Another point is that the incident flux can be used to modify the 
energy levels of the system. assuming the incident flux to be constant 
we can add a term to the potential and obtain the generalized potential
\be\label{pot2}
V(\phi)= \beta (1-\cos\phi)
+ {1\over 2} \phi^2 + \phi_e \phi,\ee
where $\phi_e $ is the incident flux, assumed constant.
This expression is plotted in 
Fig. \ref{f2a} for $\beta=15$ and $\phi_e =0, 1.8 \pi$ and $4.5 \pi$. The
minima are symmetric for $\phi_e =0$ and they are shifted to the left and the
corresponding value of the potential is decreased. By applying a sufficiently
large continuous field one can then shift the system from one state to the
other.

\begin{figure}
\centerline{\epsfig{file=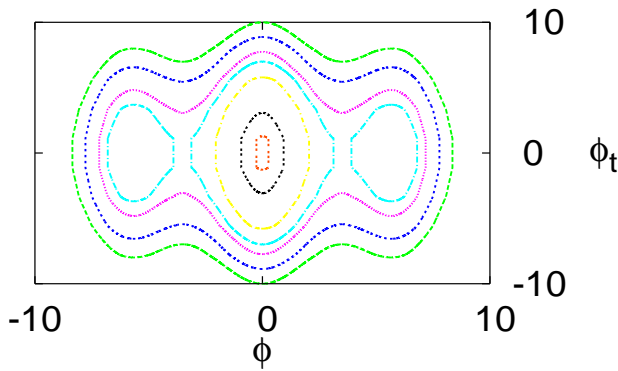,height=8 cm,width=12 cm,angle=0}}
\caption{Phase portrait $(\phi,\phi_t)$ of the Hamiltonian system
(\ref{ham}) ($\alpha=0$) for $\beta = 9.76$. The contour levels presented are
0.1, 1, 5, 17., 25.0360499136927 (separatrix) , 30.,40. and 50.}
\label{f3}
\end{figure}

For this one degree of freedom Hamiltonian, the orbits are the
contour levels of the Hamiltonian. An important orbit is the separatrix
connecting the two unstable fixed points $\phi^*\approx \pi$. It is 
given by 
\be\label{separatrix}
{1\over 2} \phi_t^2 +
\beta (1-\cos\phi)
+ {1\over 2} \phi^2 = H(\phi^*,0).\ee
This value of the Hamiltonian can be approximated for $\beta >>1$ as
\be\label{ham_separatrix}
H(\phi^*,0) \approx {\pi^2 \over 2  }(1 + {2 \over \beta  })
+ \beta(1+ \cos {\pi \over \beta})\approx {\pi^2 \over 2} + 2 \beta 
+{\pi^2 \over 2 \beta}.\ee
Fig. \ref{f3} shows the
phase portrait for $\beta = 9.76$.
For this value there are only
five fixed points. Notice the closed orbits around the
fixed points, the closed orbits surrounding the three stable fixed points.

We have shown that the steady states of the split-ring Josephson 
oscillator are similar to the stationary states of atoms. The
values $V(\phi)$ near $2n\pi$ are the analog of atomic energy levels.
In the quantum regime we should observe Metastability of these states, 
i.e. there should be quantum tunneling through the barriers at the
$ n \pi$ positions. This would result in a finite life time of these
steady states.	
In the next section, we will select the incident flux 
$\phi_{e}$ to move the system from one equilibrium to another.

\section{Influence of damping }

We consider now that the state of the system can be shifted 
from one fixed point to another via an incident flux.
For a short lived perturbation, the system then relaxes freely
to a minimum of energy. The influence of the damping is essential
, it should be present to allow the relaxation but small to
preserve the picture of the potential.
To examine how an incident flux will shift the system
from one equilibrium position to another it is useful to analyze
the work equation. To obtain it, we multiply  (\ref{srr_jj}) by
$\phi_t$ and integrate over time.
We get the difference in energy
\be\label{work_eq}
E(t_2)-E(t_1) \equiv
[{1\over 2} \phi_t^2 + \beta (1-\cos\phi) + {1\over 2} \phi^2]_{t_1}^{t_2}
= \int_{t_1}^{t_2} dt \phi_{e} \phi_t - \alpha \int_{t_1}^{t_2} dt \phi_t^2 .
\ee
The first term on the right hand side is the forcing while the second one
is the damping term. When a square pulse is applied to the system,
such that
$$\phi_{e}(t) = a ,~ {\rm for} ~ t_1 < t < t_2, ~~~0~~{\rm elsewhere}$$
the first integral is $a[\phi(t_2)-\phi(t_1)]$.
If the system is started at $(0,0)$ in phase space so that the $E(t_1)=0$
and $\phi(t_1)=0$, we have
$$E(t_2) = a\phi(t_2) - \alpha \int_{t_1}^{t_2} dt  \phi_t^2,$$
so that $\phi(t_2)$ determines how much energy is fed into the system.
When the pulse is long $t_2 >> t_1$ 
$\phi(t)$ will relax and oscillate so 
that there are values of $t_2$
such that $\phi(t_2) ~0$. In that case no energy gets fed into the system.
A sure way to avoid this is to take a narrow pulse.

The natural frequency of the oscillator around the $(0,0)$
fixed point is 
\be\label{nat_freq}
\omega_0 = \sqrt { \beta +1},\ee
which for $\beta =9.76$ gives $\omega_0\approx 3.28$ and 
a period $T_0 = 2\pi /\omega_0 \approx 1.91$. To simplify matters
we now
consider a pulse of duration much smaller than $T_0$.
This is experimentally feasible and can be modeled using
a Dirac delta function
$\phi_{e}(t)= a \delta(t)$, where $a$ is a parameter.
Let us analyze briefly the solution.
The equation (\ref{srr_jj}) becomes
\be\label{srr_jj_delta}
\phi_{tt} + (\alpha +\delta)\phi_{t} + \beta
\sin(\phi )+\phi= a \delta(t).\ee
Integrating the equation on a small interval of size $\epsilon$ around
0, we get
\be\label{jump}
[\phi_{t}]_{-\epsilon}^{\epsilon}
+ \alpha [\phi]_{-\epsilon}^{\epsilon}
+ \int_{-\epsilon}^{\epsilon} dt  (\beta \sin \phi + \phi) = a .\ee
We now take the limit $\epsilon \to 0$.
We will assume continuity of the phase so
that $[\phi]_{-\epsilon}^{\epsilon} \to 0$. The third term
being the integral of a continuous function tends to 0 when the bounds
tend to 0. Assuming $\phi_{t}(0_-)=0$ we get 
$\phi_{t}(0_+)=a$ so that such a short incident pulse will just
give momentum to the system.

We will now explore systematically the
plane $( \alpha, a)$ characteristic of the incident pulse.
The equation (\ref{srr_jj}) has been solved numerically using
a Runge-Kutta algorithm with step correction of order 4 and 5.
\begin{figure}
\centerline{\epsfig{file=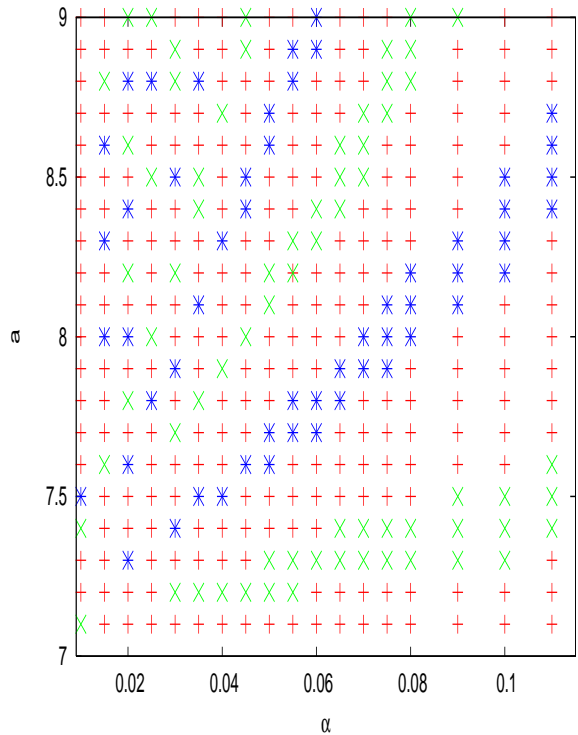,height=10  cm,width=8 cm,angle=0}}
\caption{Parameter plane $ (\alpha, a)$ showing the different 
final states reached by the system, the left focus $*$ (blue online), the 
center focus $+$ (red online) and the right focus $\times$ (green 
online). The parameter $\beta = 9.76$.}
\label{f4}
\end{figure}

%\subsection{Analysis of the trapping: probabilistic approach }

The plot in the $ (\alpha, a)$ parameter plane shown in Fig. \ref{f4}
shows the final states, $O$ the central focus ($+$), $R$ the
right focus ($\times$) and $L$ the left focus $*$ 
reached by the system. Notice how these
are organized in "tongues" following the sequence $O~R~O~L~O~R~O~L \dots$
as one sweeps the plane counterclock-wise starting from the horizontal axis.
This simple geometrical picture can be understood by
examining Fig. \ref{f3}. In the case of small damping, the separatrices
around the fixed points are not affected very much. 
Their perimeter is 
proportional to the probability of reaching one fixed point or another. 
The system is moving clock-wise along the orbits.
Assume the system reaches the central point $O$ for a given set of parameters.
If the damping is increased, the orbit might not reach $O$ but will
settle in $L$. Similarly if more kinetic energy is given to the
oscillator, it might reach $R$ instead of $O$.

Another important point is that the impulse given to the resonator must
be very short so that it relaxes following a free dynamics. The typical
frequencies of these devices are about 500 GHz so the impulse must be
around 5 Thz which is close to the frequency provided by a laser. This
seems to indicate that an optical pulse generated by a laser would be
the optimal candidate to switch the device.

To prepare the artificial in a given state, one needs to know if this state
is really reached. For that one can use
the pump-probe approach: a first pulse is sent to shift
the system in the right state, then a small second 
pulse is sent to analyze the state by reflection or
transmission. This is the object of the next section.

\section{Microwave spectroscopy of the split-ring resonator}

We consider here that the split-ring Josephson resonator is subject
to irradiation by a microwave field and compute using a scattering theory
formalism the response of the system. This field could be
microwave radiation from a wave-guide or it could be 
a laser beam shining on the device. 
The equations describing the system light-ring are the 
the generalized pendulum equation for the flux 
(\ref{usrr_jj}) and the
the Maxwell equation for the electromagnetic field  
\be\label{maxwell}
\nabla\times {\bf E} = -{\bf H}_t -{\bf M}_t,~~\nabla\times {\bf H} = {\bf E}_t,\ee
where ${\bf E}$ is the electric field, ${\bf H}$ the magnetic field and 
${\bf M}$
the magnetization. Taking the curl of the second equation we get
the vector wave equation
\be\label{vector_wave}
\nabla\times(\nabla\times {\bf H})={\bf H}_{tt} +{\bf M}_{tt}.\ee
If we assume that the wave propagates along $z$ and is transversely 
polarized so that
${\bf H}$ is parallel to $x$, the normal to the plane of the split ring, and 
${\bf E}$ is parallel to $y$.
Then we get the scalar wave equation for $H$
\be\label{scalar_wave} \Delta H - H_{tt} = M_{tt}.\ee
The magnetization is related to the current $I$ in the loop by
\be\label{mag_cur}
M = S I,\ee
where $S$ is the surface enclosed by the ring.
Combining (\ref{scalar_wave}) with (\ref{mag_cur}) and recalling the
expression of the current $I$ given by the term in brackets in
(\ref{usrr_jj})  we get the final system of equations
\begin{eqnarray}
&& H_{zz} -{1 \over c^2}H_{tt} = 
-{1\over c^2}\left ( {-\Phi_{tt} \over S} + H_{tt} \right ) l \delta(z)
,  \notag \\
[-1.5ex]  \label{h_phi} \\
&& L( C \Phi_{tt} + {\Phi_{t} \over R} +
I_c \sin({\Phi\over \phi_0}) )+\Phi=H S , \notag
\end{eqnarray}
where $l$ is the film thickness.

We introduce the units of flux, magnetic field, time and space
\be\label{units_fht}
\phi_0 = {\hbar \over 2 e},~~H_0= { \phi_0 \over S}
,~~\omega_T = {1\over \sqrt{LC}},~~ l_0 = {c\over \omega_T}.
\ee
With these units we normalize time, space, the phase and the field as
\be\label{norm_tspf}
\tau = \omega_T t,~~\zeta = {z\over l_0},~~\tilde \Phi = {\Phi \over \phi_0}
,~~\tilde H = {H\over H_0}.\ee
The normalized system obtained from (\ref{h_phi}) is then
\begin{eqnarray}
&& \tilde H_{\zeta\zeta} -\tilde H_{\tau\tau} =
-\gamma \left ( -\tilde \Phi_{\tau\tau} + \tilde H_{\tau\tau} \right )  \delta(\zeta)
,  \notag \\
[-1.5ex]  \label{h_phi_n} \\
&& \tilde \Phi_{\tau\tau} + \alpha \tilde \Phi_{\tau} + \beta
\sin(\tilde \Phi )+\tilde \Phi=\tilde H, \notag
\end{eqnarray}
where we have introduced
\be\label{abg}
\alpha = {\omega_T L \over R},~~\beta = {L I_c \over \phi_0}
,~~\gamma = { l \over l_0}.\ee

We assume that the ring is submitted to a fixed magnetic field $h_s$
to which it responds with a constant flux $\phi_s$. 
Then we send in
a small electromagnetic pulse $\delta \tilde H$ and examine the response 
$\delta \tilde \Phi$ of the ring using the scattering theory. The linearized 
equations for $\delta \tilde H,~\delta \tilde \Phi$ read
\begin{eqnarray}
&& \delta \tilde H_{\zeta\zeta} -\delta \tilde H_{\tau\tau} =
-\gamma \left ( -\delta \tilde \Phi_{\tau\tau} 
+ \delta \tilde H_{\tau\tau} \right )  \delta(\zeta)
,  \notag \\
[-1.5ex]  \label{dh_dphi} \\
&& \delta \tilde \Phi_{\tau\tau} + \alpha \delta \tilde \Phi_{\tau} + \beta
\cos(\phi_s )\delta \tilde \Phi +\delta \tilde \Phi=\delta \tilde H . \notag
\end{eqnarray}
We now assume periodic solutions 
\be\label{h_phi_harm}\delta \tilde H = h e^{i \omega \tau},~\delta \tilde\Phi
=\phi e^{i \omega \tau},\ee
and obtain the reduced system
\begin{eqnarray}
&& h_{\zeta\zeta} +\omega^2 h =
\gamma \omega^2 \left ( -\phi
+ h \right )  \delta(\zeta)
,  \notag \\
[-1.5ex]  \label{h_phi_z} \\
&& \left [-\omega^2 + i \alpha \omega +1 + \beta 
\cos(\phi_s )\right ]\phi =h . \notag
\end{eqnarray}
In the scattering we assume the electromagnetic wave to be incident
from the left of the film located at $\zeta=0$. We then have
\be\label{scat} h =  e^{-i \omega \zeta} + R e^{i\omega \zeta} ~, \zeta<0~;~~~~ 
h = Te^{-i \omega \zeta }, \zeta>0~~,\ee where 
$R$ is the amplitude of the reflected wave and
$T$ the amplitude of the transmitted wave. 
We have the following
interface conditions at $\zeta=0$ 
\be\label{interface} h(0^-)= h(0^+),~~[h_\zeta]_{0^-}^{0^+} =
\omega^2 \gamma \left ( -\phi(0) + h(0) \right ) .\ee 
They imply the two equations for $R$ and $T$
{\small \begin{eqnarray} 1+R & =& T, \notag \\
-T -(-1+R) & =& -i \omega \gamma T
\left [ {-1 \over -\omega^2 + i \alpha \omega +1 + \beta
\cos(\phi_s ) } +1 \right ] ,\notag  \end{eqnarray} }
from which we obtain the transmission coefficient,
\be\label{transmission}
T = { 2(-\omega^2 + 1 +\beta \cos \phi_s + i \omega \alpha )   
\over D},\ee
the reflection coefficient
\be\label{trans}
R = { -\alpha \gamma\omega^2 + i\gamma\omega (\beta \cos \phi_s - \omega^2)
\over D},\ee
and where the denominator is 
\be\label{den_trans_ref}
D = 2(-\omega^2 + 1+ \beta \cos \phi_s ) + \alpha \gamma\omega^2
+ i [2\alpha\omega - \gamma\omega(\beta \cos \phi_s-\omega^2)]
.\ee
\begin{figure}
\centerline{\epsfig{file=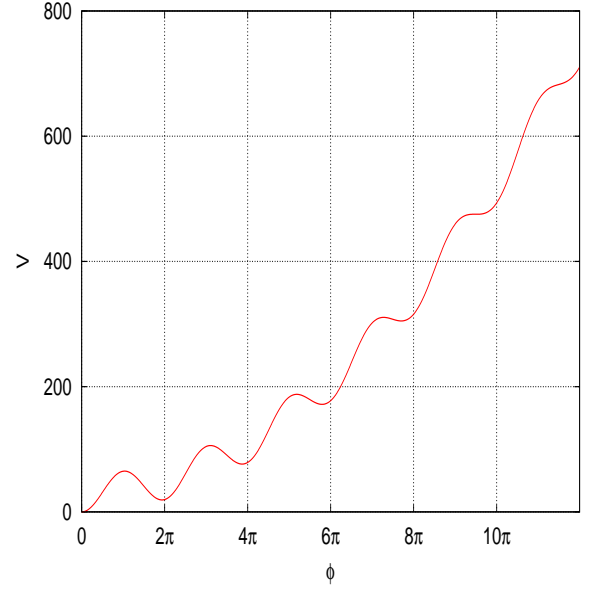,height=8  cm,width=8 cm,angle=0}}
\caption{Plot of the potential $V(\phi)$ for $\beta=30$.}
\label{f5a}
\end{figure}
\begin{figure}
\centerline{\epsfig{file=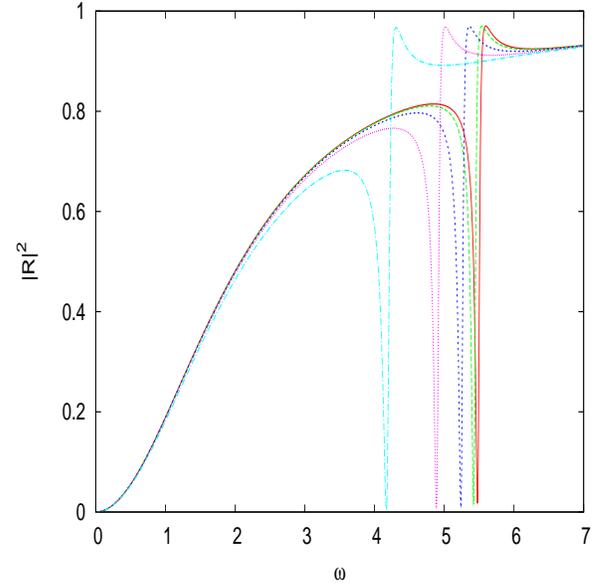,height=8  cm,width=8 cm,angle=0}}
\caption{Square of the modulus of the reflection coefficient $|R|^2$
as a function of the frequency $\omega$ for the five different equilibria
$\phi_s=0$ continuous line (red online) and $\phi_s=2 \pi (1-1/\beta)$
dashed line (green online). The parameters are $\beta=9.76,~~\gamma=1$
and $\alpha=0.01$.}
\label{f5b}
\end{figure}
The square of the modulus of $R$ is
\be\label{ref2} |R|^2 = 
{ \alpha^2 \gamma^2\omega^4+ \gamma^2\omega^2(\beta \cos \phi_s - \omega^2)^2 
\over |D|^2} .\ee
As seen in section 3, the split-ring oscillator has a finite number
of equilibria $\phi_s$ depending on the parameter $\beta$. As an
example we consider $\beta=30$ for which the potential 
$V(\phi)$ is shown in Fig. \ref{f5a}. 
The square of the modulus of the reflection coefficient (\ref{ref2})
is plotted in Fig. \ref{f5b} for the five different equilibria.
For $\omega \to 0 ~~|R|^2 \to 0$ as $\omega^2$ , for $\omega \to \infty$
$|R|^2 \to 1$. At some $\omega_s$ the transmission goes to 0,
i.e. the medium
becomes transparent. Notice the difference with a real atom which would 
absorb incident radiation for certain frequencies. 
The expression for these resonant
frequencies $\omega_s$ can be obtained by considering the
minima of $|R|^2$. These correspond to the second term in the
numerator of (\ref{ref2}) being zero. We get
\be\label{zeroesofr}
\omega_s = \sqrt{\beta\cos\phi_s} .\ee
In the example shown, the spectroscopy data $(\omega_s,\phi_s)$ is given 
in table 1.
\begin{table} \label{tab1}
\begin{tabular}
{| c | c | c | c | c | c |}
  \hline
$\omega_s$ & 5.477 & 5.420 & 5.238 & 4.888 & 4.169   \\\hline 
$\phi_s$   &  0    &  6.08 &  12.15&  18.2 & 24.18 \\
     \hline
 \end{tabular}
\caption{Spectroscopy data $(\omega_s,\phi_s)$ for a split-ring
oscillator with five steady states. The parameters are  $\beta=30,~
\gamma=1,~\alpha =0.01$.}
\end{table}

\section{Conclusion }

We have derived and analyzed a model for split ring 
Josephson resonator or RF SQUID in the superconducting regime. 
If the parameters of the device are chosen appropriately, there 
exist excited states whose number can be controlled by
carefully tuning the inductance and capacity of the ring.
We assumed that there are just two excited states and 
showed how an incident magnetic flux can shift the system from
the ground state to one of these excited states.
The existence of these excited states makes this system similar to
an artificial atom with discrete energy levels. The Josephson
oscillator is a unique nonlinear element which allows this.
Other nonlinear elements like a diode
\cite{Lapin:2003} , a Kerr material \cite{Zharov:2003} 
or a laser amplifier \cite{Gabitov_Kennedy:2010} 
would not give these discrete levels. In addition, since the
oscillator is operating in the superconducting regime the losses
are very small as opposed to the current meta-materials.

By sending a microwave field on the resonator we can perform 
a spectroscopy of it and characterize in which state it is.
Using a scattering theory formalism we compute the reflection
and transmission coefficients for the wave. These coefficients
differ clearly whether the system is in the ground state or in an
excited state enabling to distinguish them.

{\bf Acknowledgements} \\
The authors thank Matteo Cirillo and Alexei Ustinov for very helpful
discussions. JGC and AM thank the University of Arizona for 
its support. AIM is grateful to the
Laboratoire de Math\'ematiques, INSA de Rouen for hospitality and
support. The computations were done at the Centre de Ressources 
Informatiques de Haute-Normandie. This research was supported by RFBR grants No.
09-02-00701-a.

\end{document}